\documentclass[twocolumn,nofootinbib,amsmath,amssymb,aps,prd,balancelastpage,superscriptaddress]{revtex4-1}

\usepackage{color}
\usepackage[dvipsnames]{xcolor}
\usepackage[active]{srcltx}
\usepackage{amsmath,amsfonts,amssymb,amsthm,amstext,amscd,eucal,srcltx}
\usepackage{epsfig,graphicx,bm}
\usepackage{epstopdf, epsf}
\usepackage{dcolumn}
\usepackage{hyperref}
\usepackage{tensor}
\usepackage{lipsum}
\usepackage{appendix}
\usepackage{tikz}
\usepackage{caption}
\usepackage{subcaption}

\newcommand{\be}{\begin{equation}}
\newcommand{\ee}{\end{equation}}

\newcommand{\bse}{\begin{subequations}}
\newcommand{\ese}{\end{subequations}}
\newcommand{\bea}{\begin{eqnarray}}
\newcommand{\eea}{\end{eqnarray}}
\newcommand{\ba}{\begin{array}}
\newcommand{\ea}{\end{array}}
\newcommand{\bc}{\begin{center}}
\newcommand{\ec}{\end{center}}

\begin{document}
%\preprint{IPM/P-2012/009}
\vspace*{3mm}

%%%%%%%%%%%%%%%%%%%%%%%%%%%%%%%%%%%%%%%%%%%%%%%%%%%%%%%%%%%%%%%%%%%%%%%%%%%%%%%
\title{CDF II $W$-mass anomaly faces first-order electroweak phase transition}
%%%%%%%%%%%%%%%%%%%%%%%%%%%%%%%%%%%%%%%%%%%%%%%%%%%%%%%%%%%%%%%%%%%%%%%%%%%%%%%

\author{Andrea Addazi}
\email{addazi@scu.edu.cn}
\affiliation{Center for Theoretical Physics, College of Physics Science and Technology, Sichuan University, 610065 Chengdu, China}
\affiliation{Laboratori Nazionali di Frascati INFN, Frascati (Rome), Italy, EU}

\author{Antonino Marcian\`o}
\email{marciano@fudan.edu.cn}
\affiliation{Center for Field Theory and Particle Physics \& Department of Physics, Fudan University, 200433 Shanghai, China}
\affiliation{Laboratori Nazionali di Frascati INFN, Frascati (Rome), Italy, EU}

\author{Ant\'onio P. Morais} 
\email{aapmorais@ua.pt}
\affiliation{Theoretical Physics Department, CERN, 1211 Geneva 23, Switzerland}
\affiliation{Departamento de F\'\i sica da Universidade de Aveiro and Centre  for  Research  and  Development in  Mathematics  and  Applications (CIDMA), Campus de Santiago, 3810-183 Aveiro, Portugal, EU}

\author{Roman Pasechnik}
\email{Roman.Pasechnik@hep.lu.se}
\affiliation{Department of Physics,
Lund University, SE 223-62 Lund, Sweden, EU}

\author{Hao Yang} 
\email{hyang19@fudan.edu.cn}
\affiliation{Center for Field Theory and Particle Physics \& Department of Physics, Fudan University, 200433 Shanghai, China}

\begin{abstract}
\noindent
We suggest an appealing strategy to probe a large class of scenarios beyond the Standard Model simultaneously explaining the recent CDF II measurement of the $W$ boson mass and predicting first-order phase transitions (FOPT) testable in future gravitational-wave (GW) experiments. Our analysis deploys measurements from the GW channels and high energy particle colliders. We discuss this methodology focusing on the specific example provided by an extension of the Standard Model of particle physics that incorporates an additional scalar $\mathrm{SU}(2)_{\rm L}$ triplet coupled to the Higgs boson. We show that within this scenario a strong electroweak FOPT is naturally realised consistently with the measured $W$ boson mass-shift. Potentially observable GW signatures imply the triplet mass scale to be TeV-ish, consistently with the value preferred by the $W$ mass anomaly. This model can be tested in future space-based interferometers such as LISA, DECIGO, BBO, TianQin, TAIJI projects and in future colliders such as FCC, ILC, CEPC. 
\end{abstract}
\begin{flushright}
CERN-TH-2023-039\\
\vskip1cm
\end{flushright}

\maketitle

%%%%%%%%%%%%%%%%%%%%%%%%%%%%%%%%%%%%%
\section{Introduction}
\label{Sec:intro}
%%%%%%%%%%%%%%%%%%%%%%%%%%%%%%%%%%%%%
\noindent
The CDF II Collaboration has recently reported a new and quite unexpected result from the $W$ boson mass measurement \cite{CDF:2022hxs}, which lies $7.2\sigma$ away from theoretical predictions of the Standard Model (SM) of particle physics \cite{ParticleDataGroup:2020ssz}. In order to explain this anomaly, several scenarios beyond the SM have been recently suggested in the literature. In particular, new states have been incorporated including additional $\mathrm{SU}(2)_{\rm L}$ Higgs doublets, vector-like fermion $\mathrm{SU}(2)_{\rm L}$ triplets, vector-like top partners, leptoquarks, singlet-doublet fermion pairs, scalar $\mathrm{SU}(2)_{\rm L}$ triplets and quadruplets, right-handed neutrinos, $Z'$ and extra vector bosons, FIMP dark matter modes, $U(1)_{L_\mu-L_\tau}$ modes, vectorlike quarks, canonical scotogenic neutrino-dark matter modes, $U(1)_{L_\mu-L_\tau}$ vector-like leptons --- see e.g. Refs.~\cite{Ahn:2022xeq,Han:2022juu,Perez:2022uil,DiLuzio:2022xns,Asadi:2022xiy,Popov:2022ldh,Du:2022fqv,Balkin:2022glu,Gu:2022htv,Cheng:2022jyi,Bhaskar:2022vgk,Ghorbani:2022vtv,Borah:2022obi,Chowdhury:2022moc,Zhang:2022nnh,Kanemura:2022ahw,Ghoshal:2022vzo,Heo:2022dey,Cheung:2022zsb,Endo:2022kiw,Bagnaschi:2022whn,Bahl:2022xzi,Athron:2022isz,Blennow:2022yfm,Fan:2022dck,Strumia:2022qkt,Biekotter:2022abc,Bandyopadhyay:2020otm,Liu:2022jdq,Lee:2022nqz,Baek:2022agi,Sakurai:2022hwh,Cao:2022mif,Batra:2022pej,Zhou:2022cql}. Also a top-down motivated model has been considered, in which extra states come from a D3-brane \cite{Heckman:2022the}.
Implications for electroweak baryogenesis and Chameleon dark energy have been also considered \cite{Huang:2015izx,Yuan:2022cpw}, while the relevance of hadronic uncertainty and electroweak precision tests for the correct interpretation of the result and the prospect on new physics has been delved in \cite{Athron:2022qpo,Lu:2022bgw,Fan:2022yly}.

This large and statistically significant anomaly within the Electro-Weak (EW) sector urges us to question what are its possible implications for our understanding of the EW phase transitions (EWPTs), and more in general whether it can be related to a first-order phase transition (FOPT) in the early Universe. At the first sight, a relation between the $W$ mass anomaly and the order of cosmological phase transitions may appear not so direct and clear. Certainly, the answer would be model-dependent. 

In this short letter, we do not pretend to be exhaustive in covering the wealth of phenomenologically allowed models that address this broad research topic. We rather seek to answer questions related to the aforementioned relevant issues, focusing on a specific simplified framework that relates the parameter space of EWPTs to a possible explanation of the $M_W$-anomaly. Specifically, the model we consider is based on a minimal scalar-triplet extension of the SM scalar sector providing a natural explanation of the anomaly as suggested earlier in Ref.~\cite{DiLuzio:2022xns}. Even without providing here a detailed scan of the parameter space of the considered minimal model, this simplified approach will nevertheless help us gain intuition on the way our outlined strategy for new physics search can be applied to other theoretical frameworks invoked to explain the measured $M_W$-anomaly.

We will show that, in a large subset of the parameter space of the simplified model, we obtain gravitational-wave (GW) stochastic background signals as echos of the strong EW FOPT in the early Universe that can be tested in space-based interferometers such as LISA, DECIGO, BBO, TianQing and TAIJI to be deployed in foreseeable future. We also notice that the considered parameter space of the model does not violate the current LHC constraints. The model can also be tested at future linear or circular lepton colliders such as the FCC, ILC and CEPC, through a measurement of the trilinear Higgs coupling, which receives relatively large corrections due to Higgs interactions with the scalar triplet.

%%%%%%%%%%%%%%%%%%%%%%%%%%%%%%%%%%%%%%%%%%%%%%%%%%%%%%%%%%%%%%%%%%%%%%%%%%%%%%%%%%%%%
\section{Minimal $\mathrm{SU}_{\rm L}(2)$ triplet extension: $M_W$-anomaly and FOPT}
\label{Sec:model}
%%%%%%%%%%%%%%%%%%%%%%%%%%%%%%%%%%%%%%%%%%%%%%%%%%%%%%%%%%%%%%%%%%%%%%%%%%%%%%%%%%%%%
\noindent
The CDF-II measurement of the $W$ boson mass $M_W$ suggests an anomaly in the $\hat{T}$-parameter \cite{Strumia:2022qkt} (in particular, under an assumption of $\hat{U}=0$), namely 
\begin{eqnarray} 
\label{TKK}
&\hat{T}\simeq (0.84\pm 0.14)\times 10^{-3}\,, \\
&c_{\rm HD}=-(0.17\pm 0.07/{\rm TeV})^{2}\, ,
\end{eqnarray}
with $c_{\rm HD}$ being the coupling related to the Effective Field Theory (EFT) operator expressed by
\begin{equation}
\label{OHD}
c_{\rm HD}\mathcal{O}_{\rm HD}=c_{\rm HD}(H^{\dagger}D_{\mu}H)((D_{\mu}H)^{\dagger}H)\, ,
\end{equation}
and the $\hat{T}$-parameter by
\begin{equation}
\label{TTT}
\hat{T}=-\frac{v^2}{2}c_{\rm HD}\, .
\end{equation}

A possible simple explanation that has been suggested is to introduce a new state, $\Delta=(1,3,0)$ of mass $M_{\Delta}$, with charges given w.r.t.~the SM gauge group $\mathrm{SU}(3)_{\rm c}\times \mathrm{SU}(2)_{\rm L} \times \mathrm{U}(1)_{\rm Y}$, i.e.~a real scalar triplet of $\mathrm{SU}(2)_{\rm L}$ that is a color singlet and has no hypercharge \cite{DiLuzio:2022xns}. This state is coupled to the Higgs doublet via the interaction Lagrangian term 
\begin{equation}
\label{LT}
\mathcal{L}_{\rm T}=-k_{\Delta}H^{\dagger} \Delta \cdot \sigma H+ {\rm h.c.}\,, 
\end{equation}
where $\sigma$ denotes the Pauli matrices. Integrating out the massive state $\Delta$, the interaction term \eqref{LT} directly generates negative coupling in the EFT operator \ref{OHD}
\begin{equation}
\label{DEE}
c_{\rm HD}=-2\frac{k_{\Delta}^{2}}{M_{\Delta}^{4}}\, , 
\end{equation}
and hence leading to a positive $\hat{T}$ contribution consistent with the observed shift in the $W$ mass,
\begin{equation}
\label{TTA}
\hat{T}=\frac{k_{\Delta}^2 v^2}{M_{\Delta}^4}=0.84 \times 10^{-3}\left(\frac{|k_{\Delta}|}{M_{\Delta}}\right)^2 \left(\frac{8.5\, {\rm TeV}}{M_{\Delta}}\right)^2\, .
\end{equation}
This is a tree-level effect suggesting that the $\mathrm{SU}(2)_{\rm L}$ scalar triplet can be in a multi-TeV mass range. Nonetheless, saturating the perturbativity bound $|k_{\Delta}|/M_{\Delta}\leq 4\pi$, the triplet cannot exceed 100 TeV \cite{DiLuzio:2022xns}.

It is worth to notice that, after integrating out $\Delta$, Eq.~\ref{LT} generates an additional contribution to the quartic Higgs self-interaction term of the form  $(k_{\Delta}/m_{\Delta})^2 (H^{\dagger}H)^2$. The Higgs bare coupling constant $\lambda_{\rm bare}$ hence receives a tree-level correction, according to $\lambda=\lambda_{\rm bare}+(k_{\Delta}/m_{\Delta})^2$. In what follows, we consider the full Higgs quartic coupling $\lambda=m^2/2v^2$ (with $m^2$ being the Higgs mass parameter in the Lagrangian and $v\simeq 246$ GeV -- the Higgs vacuum expectation value) rather than $\lambda_{\rm bare}$, which appears in the SM framework.

We may now focus on a Lagrangian term of the form 
\begin{equation}
\label{DTTR}
\frac{\mu_{\Delta}}{3}\Delta^3 + {\rm h.c.} \,,
\end{equation}
where $\Delta^3\equiv (\Delta \cdot \sigma)(\Delta \cdot \sigma)(\Delta \cdot \sigma)$. Integrating out $\Delta$-states, Eqs.~\eqref{LT} and \eqref{DTTR} generate the following six-dimensional operator
\begin{equation}
\label{eighttt}
\frac{\kappa}{\Lambda^2}(H^{\dagger}H)^3 + {\rm h.c.}
\end{equation}
in terms of the cutoff scale $\Lambda$, where
\begin{equation}
\label{kkka}
\frac{\kappa}{\Lambda^2}=\frac{\mu_{\Delta}k_{\Delta}^{3}}{3M_{\Delta}^6}\, .
\end{equation}
The latter recasts as  
\begin{equation}
\label{Lamk}
 \frac{\Lambda}{\sqrt{\kappa}}=\frac{\sqrt{3}M_{\Delta}^3}{\sqrt{\mu_{\Delta}}k_{\Delta}^{3/2}}\, , 
 \end{equation}
with $\kappa\lesssim 4\pi$ as a perturbativity bound. Note, the six-dimensional operator \eqref{eighttt} appears to be a crucial contribution to determine the nature and the strength of the EWPT. 

In order to develop a consistent analysis of the EW FOPT, it is convenient to choose the unitary gauge, such that $H=h/\sqrt{2}$. The one-loop finite-temperature effective potential then casts as 
\begin{equation}
V_{\rm eff}(T,h)=V_{\rm tree}(h)+V_{T=0}^{(1)}(h)+\Delta V_{T}(h,T)\,,
\end{equation}
where 
\begin{equation}
V_{\rm tree}(h)=\frac{1}{2}m^2 h^2+\frac{\lambda}{4}h^4 + \frac{\kappa}{8\Lambda^2}h^6
\end{equation}
is the tree-level Higgs potential, $V_{T=0}^{(1)}(h)$ is the Coleman-Weinberg one-loop potential fixed at the EW scale at zero temperature, and $\Delta V_{T}(h,T)$ is the thermal contribution obtained through the daisy resummation technique \cite{Grojean,Delaunay:2007wb} and the use of dimensional reduction within the context of EWPT thermodynamics \cite{Schicho:2021gca,Niemi:2021qvp,Schicho:2022wty,Croon:2020cgk}.

At tree-level, the effective Higgs potential acquires a dominant thermal correction to the mass that reads as $CT^2/2$, where 
\begin{equation}
\label{CC}
C\simeq \frac{1}{16}\Big(g'^2+3g^2+4y_{\rm t}^2+4\frac{m_\mathrm{h}^2}{v^2}+36\frac{\kappa v^2}{\Lambda^2}\Big)\,,
\end{equation}
and where $g',\,g$ are, respectively, the $\mathrm{U}(1)_{\rm Y}$ and $\mathrm{SU}(2)_{\rm L}$ gauge couplings, $y_{\rm t}$ is the Yukawa coupling of the top quark providing a leading contribution from the SM fermion sector and $m_\mathrm{h}$ is the Higgs boson mass which, at tree-level, is given by $m^2_\mathrm{h} = 2 \lambda v^2 + 3 v^4 \kappa / \Lambda^2$. 
In this work, we compute the Coleman-Weinberg contribution and perform the bounce action calculations and the search for FOPTs using the \texttt{CosmoTransitions} package \cite{Wainwright:2011kj}.

As it was previously found in \cite{Grojean,Delaunay:2007wb,Huang:2016odd}, the required condition to induce strong FOPTs in effective extensions of the Higgs sector with dimension-6 operators implies that the $\Lambda/\sqrt{\kappa}$ energy scale is limited in the range $480 \div 840\, {\rm GeV}$. In this article a concrete UV realization is considered such that, using relation \eqref{Lamk}, one can recast this range in terms of the $\Delta$-sector parameters as
\begin{equation}
\label{ggghh}
 480\, {\rm GeV}\lesssim \frac{\sqrt{3}M_{\Delta}^3}{\sqrt{\mu_{\Delta}}k_{\Delta}^{3/2}}\lesssim 840\, {\rm GeV}\, ,
\end{equation}
which will be used as input in our numerical analysis. The FOPT conditions that must be satisfied are $T_{\rm c}>0$ and $v(T_{\rm c})/T_{\rm c}>1$, in terms of the critical temperature of the phase transition, $T_{\rm c}$. These lead to the range in the cutoff scale $\Lambda_{\rm m}\leq \Lambda \leq \Lambda_{\rm M}$, which in turn corresponds to the observed Higgs mass $m_{\rm h}=125\, {\rm GeV}$. This is found employing the following relations for $\lambda,\, m$ parameters in the Higgs sector: $m^2=m_{\rm SM}^{2}(1-\Lambda_{\rm M}^2/2\Lambda^2)$ and $\lambda=\lambda_{\rm SM}(1-\Lambda_{\rm M}^2/\Lambda^2)$, with $\Lambda_{\rm M}=\sqrt{3}\Lambda_{\rm m}=\sqrt{3\kappa}v^2/m^2$, and $m_{\rm SM}^{2}$, $\lambda_{\rm SM}$ being the SM counterparts.

The bounds imposed by the FOPT conditions allow for a scalar triplet to be in a multi-TeV mass range. Saturating the perturbative bounds for the triplet mass $M_{\Delta}$ as $|k_{\Delta}|/M_{\Delta}\simeq 4\pi$ and $|\mu_{\Delta}|/M_{\Delta}\simeq 4\pi$, the FOPT bounds in Eq.~\eqref{ggghh} correspond to $M_{\Delta}\simeq 5\div 10\, {\rm TeV}$.

A strong EW FOPT sources bubble nucleation via quantum tunneling and thermal fluctuations from a metastable false vacuum to the true vacuum. The dynamics of phase transitions are characterized by $T_{*},\,\alpha,\,\beta$ parameters. Here, $T_{*}$ stands for the percolation temperature, at which the probability of finding a point in the false vacuum is 0.7 \cite{Ellis:2020nnr}. The $\alpha$ parameter reads $\alpha\equiv \epsilon(T_{*})/\rho_{\rm rad}(T_{*})$, with $\epsilon(T)$ being the latent heat and $\rho_{\rm rad}(T)$ -- the primordial plasma thermal energy. The $\beta$ parameter is the characteristic time scale of the EWPT, and is related to the size $d$ of the bubble as $d\simeq v_{b}/\beta$, with $v_{b}$ being bubble wall expansion velocity. The key parameters are all controlled by the effective scalar potential according to the following relations:
\begin{equation}
\label{alphat}
\alpha=\frac{30}{\pi g_{*}(T_{*})T_{*}^4}\Big[\frac{T}{4}\frac{d \Delta V^{\rm min}_{\rm eff}(T,h)}{dT}- \Delta V^{\rm min}_{\rm eff}(T,h)\Big]_{T=T_*}\,, 
\end{equation}
\begin{equation}
\label{betat}
\beta=-\frac{dS_{\rm E}}{dt}\Big|_{t=t_{*}}\,,
\end{equation}
where $S_{\rm E}(T)$ denotes the bubble 3D Euclidean action divided by the temperature and $t_{*}$ is the cosmological time at which $T=T_{*}$, $g_{*}(T_{*})$ are the relativistic degrees of freedom at $T=T_{*}$ and $\Delta V^{\rm min}_{\rm eff}(T,h)$ represents the difference of the effective potential before and after the transition takes place at $T_\ast$.

The $T_{*},\,\alpha,\,\beta$ parameters introduced above characterize the GW energy spectrum, which receives three main contributions from bubble collisions \cite{bubble}, sound shock waves \cite{sound1} and turbulence \cite{tur1,tur2}, all described by well-known semi-analytical formulas. Simulations of FOPTs from a specific field theory provide an input for the semi-analytical formulas, which in turn generate the related characteristics of GW spectra as output. Within this analysis we deploy standard methods in accounting for collision, turbulence and sound-wave contributions --- see e.g. Ref.~\cite{Caprini:2015zlo,Caprini:2019egz}).

\begin{figure}
%\centering
\begin{center}
\includegraphics[width=1.0 \linewidth]{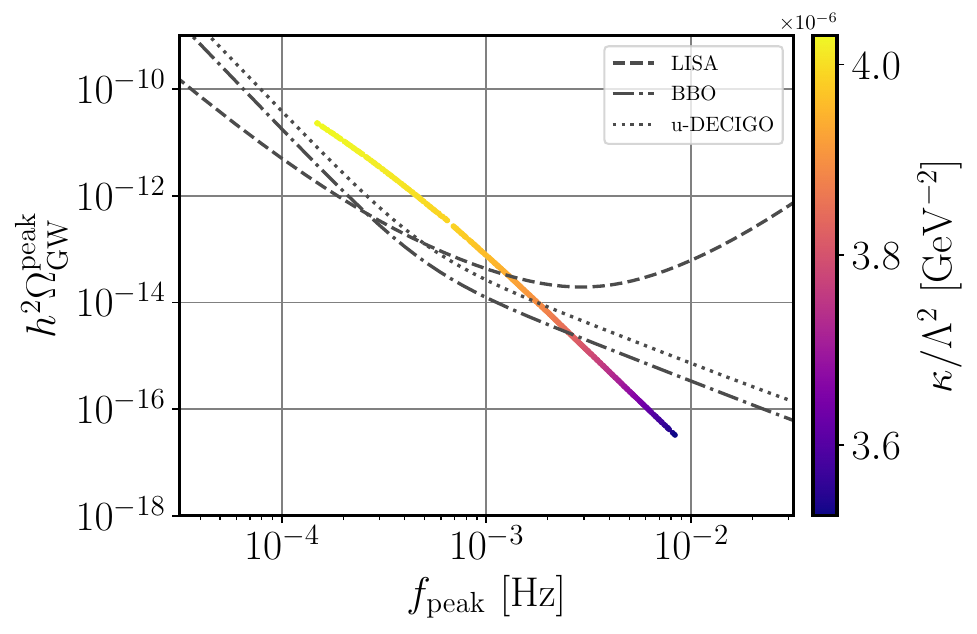}
\end{center}
\vspace{-0.6cm}
\caption{\footnotesize Parametric scan displayed in $\hat{T}$, which varies within the range $[0.76, 0.84]\times 10^{-3}$.
}
\label{fig:all}
\end{figure}

\begin{figure}
%\centering
\begin{center}
\includegraphics[width=1.0 \linewidth]{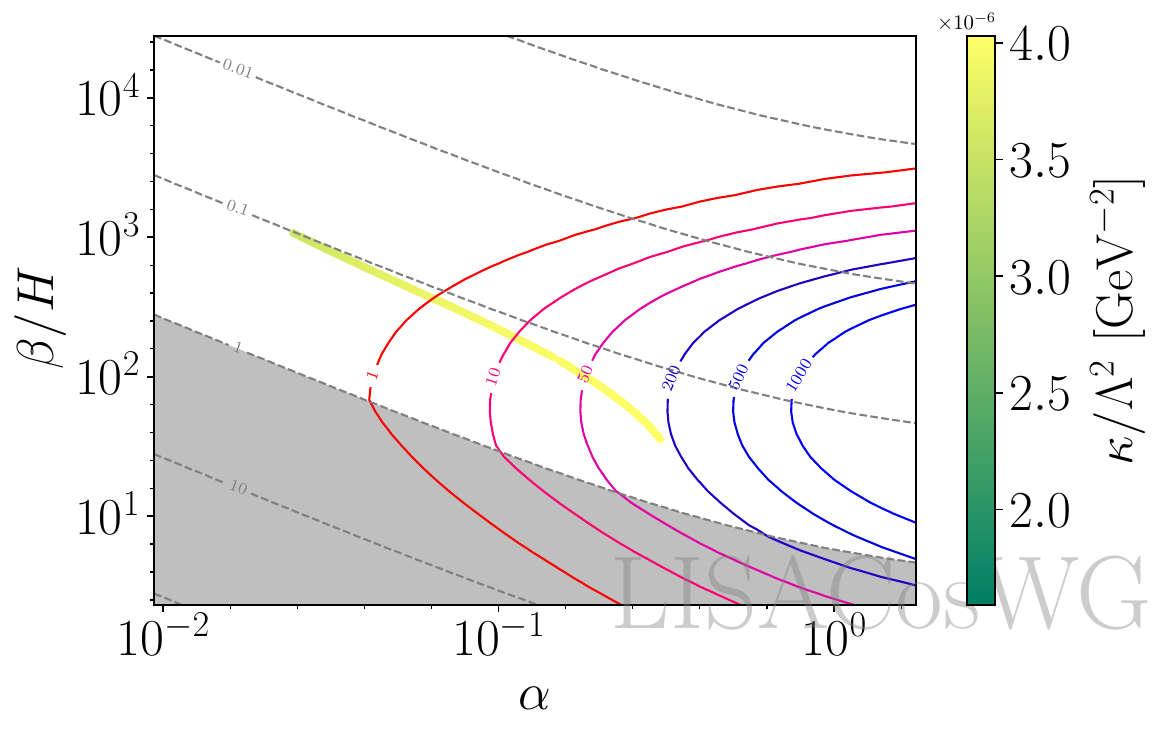}
\end{center}
\vspace{-0.6cm}        
\caption{\footnotesize 
Scatter plots displaying the SNR for the points detectable by LISA. The color bar denotes the value of $\hat{T}$. Higher values of the range of $\hat{T}$ correspond to higher intensities of the GW signals.
}
\label{fig:SNR}
\end{figure}

We have performed a parametric scan by varying $\Lambda/\sqrt{\kappa}$ in the range $[480,840]~\mathrm{GeV}$ using a numerical routine based on \texttt{CosmoTransitions} \cite{Wainwright:2011kj} to calculate the phase transition parameters $\alpha$ and $\beta$, as well as the GW's peak amplitude ($h^2 \Omega^\mathrm{peak}_\mathrm{GW}$) and frequency ($f_\mathrm{peak}$). As shown in Fig.~\ref{fig:all} one can notice that strong FOPTs associated to the production of potentially visible GWs at LISA and future interferometers restricts $\Lambda/\sqrt{\kappa}$ to a narrow region of approximately $[500,510]~\mathrm{GeV}$. Such a result is rather tantalizing, not only because it corresponds to a TeV scale triplet, but, above all, this is indeed the preferred region favoured by the CDF II W mass anomaly. In particular, expressing the parametric scan in term of $\hat{T}$, which is related\footnote{We have fixed $\mu_\Delta=7.33$ PeV, which corresponds to the value $\Lambda/\sqrt{\kappa}=500$ GeV that maximizes the amplitude of the GWs and for which the related SNR is greater than $20$, having used the experimental result for $\hat{T}$ specified by Eq.~\eqref{TKK}.} to $\Lambda/\sqrt{\kappa}$ through Eq.~\eqref{TTA}, we have found that higher values of the parameter $\hat{T}$, related to lower values of the energy range of $\Lambda/\sqrt{\kappa}$, correspond to higher intensities of the GWs stochastic background that would be originated, as Fig.~\ref{fig:all} and Fig.~\ref{fig:SNR} clearly depict. Therefore, to higher values of the parameter $\hat{T}$ measured by CDF II correspond higher values of the amplitude of the GWs signal and of the related signal-to-noise ratio (SNR).  

Varying $\hat{T}$ within the same range $[0.76, 0.84]\times 10^{-3}$, we can show in Fig.~\ref{fig:SNR} the SNR that corresponds to points detectable by LISA. In particular, for SNR greater than 20, one obtains $\hat{T}=0.844 \times 10^{-3}$, $e.g.$ the first point in Tab.~\ref{table-example}, which corresponds to $\Lambda/\sqrt{\kappa} \sim 500~ \mathrm{GeV}$.

\begin{figure}%[H]
%\centering
\begin{center}
\includegraphics[width=1.00 \linewidth]{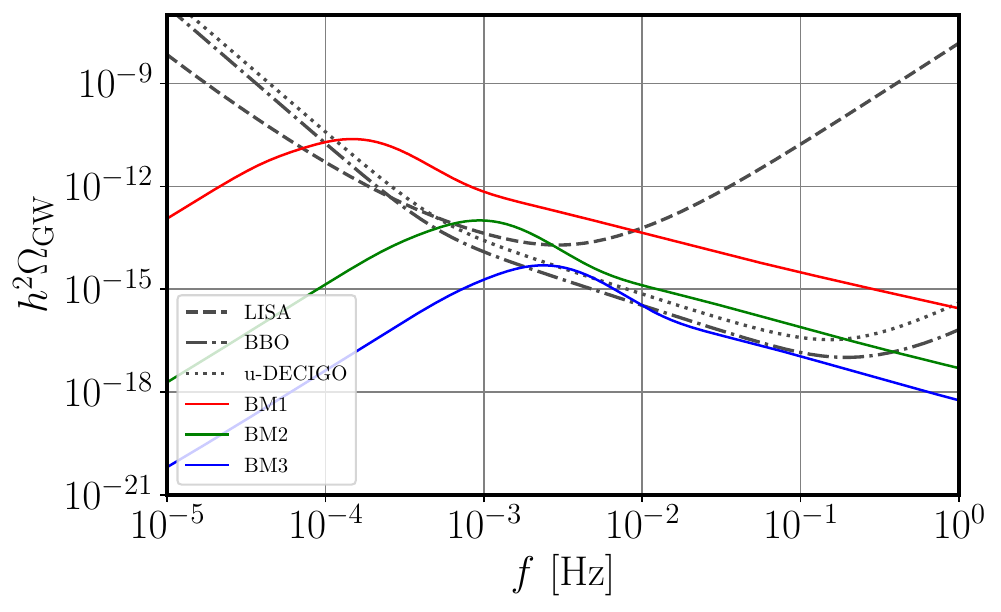}
\end{center}
\vspace{-0.6cm}        
\caption{\footnotesize The GW spectra for the three benchmark FOPTS solution listed in Table I have been plotted in a frequency domain that allows to make comparisons with the sensitivity curves of LISA, BBO and u-DECIGO. }
\label{fig:benchmark}
\end{figure}

\begin{table}[h]
\centering
\scalebox{1.3}{
\begin{tabular}{|c|c|c|c|c|}
    \hline
$T_{*}(\mathrm{GeV})$ & $\alpha$ & $\beta/H_{*}$ & $\hat{T}$ & $\delta_{\sigma_{hz}}(\%)$ \\
    \hline
    $43.8$ & $0.30$ & $36.37$ & $0.844 \times 10^{-3}$ & $3.02$ \\
   \hline
   $55.6$ & $0.12$ & $180.94$ & $0.835 \times 10^{-3}$ & $2.97$\\
   \hline
    $64.2$ & $0.07$ & $394.14$ & $0.822 \times 10^{-3}$ & $2.90$\\
    \hline
\end{tabular}
}
\caption{Benchmark FOPT solutions that can be detected in future GW space-based interferometers. }
\label{table-example}
\end{table}

\begin{figure}%[H]
%\centering
\begin{center}
\includegraphics[width=0.85\linewidth]{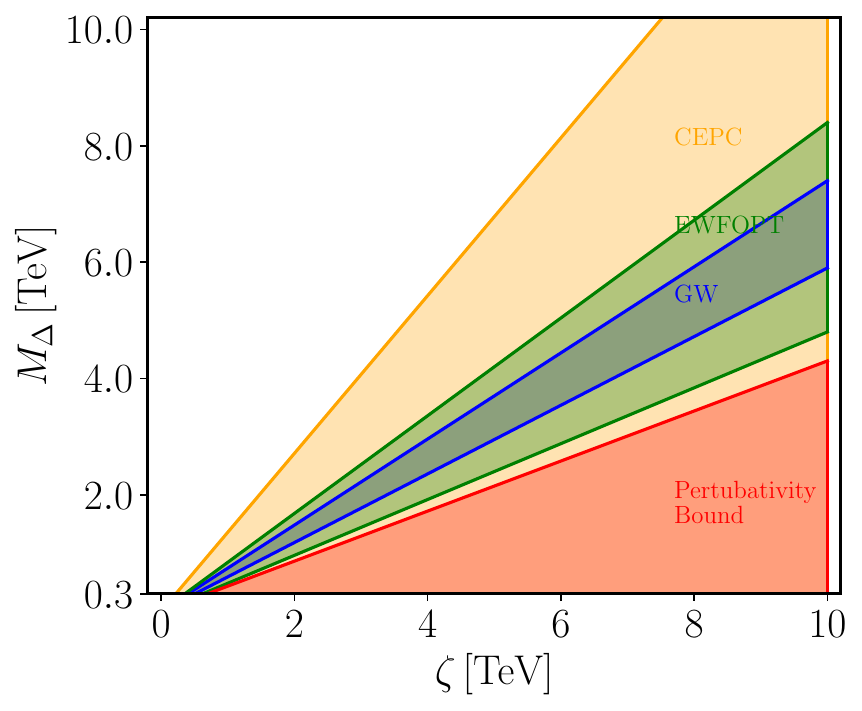}
\end{center}
\vspace{-0.6cm}        
\caption{\footnotesize Parametric regions of the mass of the triplet $M_{\Delta}[{\rm TeV}]$ and $\zeta[{\rm TeV}]=(\mu_{\Delta}/3)^{1/3}k_{\Delta}^{2/3}$ allowing FOPT (Green) and GW detectable in LISA, u-DECIGO, BBO (Blue), in comparison with CEPC test capability (orange). The exclusion region from the perturbativity bound for $\Delta$-couplings is displayed (Red). }
\vspace{-0.6cm}   
\label{fig:viable-regions}
\end{figure}

In Tab.~\ref{table-example}, we have listed three scenarios corresponding to the generation of EWPT in the model under scrutiny. We show that these FOPT branches can be promisingly tested in space-based interferometers (see Fig.~\ref{fig:benchmark}). As we expected, for the three cases corresponding to the benchmarks in Tab.~\ref{table-example}, we find that non-runaway bubble solutions and sound shock wave and turbulence contributions are predominant with respect to bubbles' collision ones. We decided to focus on these three examples, since not only they evade LHC bounds on direct searches and trilinear Higgs coupling, but they can also be tested 
at CEPC. Indeed, in the model we are considering the Higgs trilinear coupling $\lambda_{3h}$ is expressed by
\begin{equation}
\lambda_{3h}=-(1+\delta_h)\frac{A h^3}{6}\,,
\end{equation}
where $A=3m_{h}^2/v$ and $\delta_h=2\Lambda_m/\Lambda$. Here  $\delta_h$ varies within the range $0.66 \div 2$, the values of which can be compared with the $hZ$ cross section data $\sigma_{hZ}$, with precision $\delta_{\sigma_{hZ}}=\delta\sigma_{hZ}/\sigma_{hZ}$.
CEPC can achieve the precision $\delta_{\sigma_{hZ}}\simeq 1.6\%$ at $\sqrt{s}=240\, {\rm GeV}$ collision energy, corresponding to $\delta_h(\kappa=1)=0.25$ for integrated luminosity of $10\,ab^{-1}$ --- see e.g.~Refs.\cite{CEPC,Huang:2016odd}. Thus CEPC can directly probe the model we considered testing both EWPT and $M_{W}$-anomaly from heavy scalar triplet --- see Fig.~\ref{fig:benchmark} and Fig.~\ref{fig:viable-regions}.

Note that the measurement of the triple Higgs coupling can achieve the statistical significance of $4.5\sigma$ at a potential high-energy 27 TeV LHC (HE-LHC) upgrade \cite{Homiller:2018dgu}. This, together with the observed $W$-mass anomaly and a possible primordial GWs detection, offers a striking opportunity for probing the considered triplet extension of the SM in a not too distant future.

%%%%%%%%%%%%%%%%%%%%%%%%%%%
\section{Conclusions}
\label{Sec:conclusion}
%%%%%%%%%%%%%%%%%%%%%%%%%%%
\noindent 
In this Letter, we have explored an interplay between the recently observed anomaly detected in the $W$-mass measurement by the CDF-II Collaboration and the dynamics of the strong first-order Electro-Weak (EW) phase transitions. For this purpose, we have considered an insightful example of a model for new physics containing a scalar $\mathrm{SU}(2)_{\rm L}$ triplet with only three adjustable free parameters on top of those of the Standard Model (SM): a triplet mass term and its trilinear self-coupling as well as a trilinear coupling to the Higgs boson. We have found that even in this simplified framework one can naturally explain the observed new physics correction to the $W$ mass while sourcing a strong first-order EW phase transition that potentially generates observable primordial gravitational wave (GW) signatures in cosmology. 

The considered minimal $\mathrm{SU}(2)_{\rm L}$ triplet extension of the SM is an important example of more extended Beyond SM scenarios that predict a sizeable dimension-6 $(HH^\dagger)^3$ operator in the Higgs sector above the EW scale leading to first-order phase transitions in the EW sector. 
%and, simultaneously, enables to describe the CDF-II $W$ mass anomaly. 
Such a heavy triplet emerges, for instance, in the context of $\mathrm{SU}(5)$ Grand-unified field theory \cite{FileviezPerez:2008bj,Strumia:2022qkt,FileviezPerez:2022lxp,Senjanovic:2022zwy}, as a scalar adjoint representation of dimension $24$.
Our analysis shows that the existence of potentially observable GW signatures implies the triplet mass scale to be TeV-ish, which in turn is close to the value preferred by the $W$ mass anomaly. With this example, our analysis explicitly demonstrates that a class of models featuring a $\mathrm{SU}(2)_{\rm L}$ triplet scalar state can be probed by future GWs interferometers such as LISA, DECIGO, TianQin and TAIJI, around the mHZ frequency scale, as well as from measurements of the trilinear Higgs coupling in future linear or circular colliders. 

\vspace{0.3cm}
%%%%%%%%%%%%%%%%%%%%%%%%%%
{\bf Acknowledgements}\\
%%%%%%%%%%%%%%%%%%%%%%%%%%
\noindent 
A.A.~work is supported by the Talent Scientific Research Program of College of Physics, Sichuan University, Grant No.1082204112427 \& the Fostering Program in Disciplines Possessing Novel Features for Natural Science of Sichuan University,  Grant No. 2020SCUNL209 \& 1000 Talent program of Sichuan province 2021. 
A.M.~wishes to acknowledge support by the Shanghai Municipality, through the grant No.~KBH1512299, by Fudan University, through the grant No.~JJH1512105, the Natural Science Foundation of China, through the grant No.~11875113, and by the Department of Physics at Fudan University, through the grant No.~IDH1512092/001.
A.P.M~is supported by national funds (OE), through FCT, I.P., in the scope of the framework contract foreseen in the numbers 4, 5 and 6 of the article 23, of the Decree-Law 57/2016, of August 29, changed by Law 57/2017 of July 19 and by the projects PTDC/FIS-PAR/31000/2017, CERN/FIS-PAR/0014/2019, CERN/FIS-PAR/0027/2019, UIDB/04106/2020 and UIDP/04106/2020.
R.P.~is supported in part by the Swedish Research Council grant, contract number 2016-05996, as well as by the European Research Council (ERC) under the European Union's Horizon 2020 research and innovation programme (grant agreement No 668679).
The authors would like to thank João Viana for his valuable contribution in the improvement of the routines used to determine the GW spectra, in particular, for building a function to smooth the bounce action in \texttt{CosmoTransitions} and that was first presented in \cite{Freitas:2021yng}. The authors are also thankful to Felipe F.~Freitas for having adapted the \texttt{PTPlot} tool \cite{Caprini:2019egz} with the inclusion of a colour scale feature in SNR plots.

%%%%%%%%%%%%%%%%%%%%%%%%%%%%%%%%%%%%

\end{document}